# LA EXPLICACIÓN DE LOS ECLIPSES EN LA ANTIGÜEDAD GRECOLATINA

**Roberto Casazza**
Universidad Nacional de Rosario
Universidad de Buenos Aires
casazza.roberto@gmail.com

**Alejandro Gangui**
Instituto de Astronomía y Física del Espacio (CONICET)
Universidad de Buenos Aires
relat@iafe.uba.ar

**Resumen:** La búsqueda de una comprensión racional de los eclipses domina los albores del pensamiento filosófico y científico. En ese marco, el pleno conocimiento del "ciclo de saros" (ciclo de 18 años, 10 u 11 días y 1/3 de día que establece el período de tiempo que media entre dos eclipses totales de sol o de luna sucesivos con muy similares características), heredado por los griegos de los babilonios, permitió importantes desarrollos teóricos en el ámbito grecolatino. Nuestro propósito en este trabajo será: a) indagar, sintética y esquemáticamente, las teorizaciones antiguas sobre las causas de los eclipses, b) comentar acerca de la predictibilidad de los eclipses en la Antigüedad, c) describir algunas recomendaciones –presentes en textos clásicos– para la observación de eclipses de sol, y d) reseñar sucintamente las creencias populares y prácticas sociales ante la ocurrencia inesperada de los eclipses.

**Palabras clave:** eclipses – astronomía – saros – cosmología

**Abstract:** The search of a rational explanation of eclipses pervades the beginnings of philosophical and scientific thought. Within this intellectual frame, the knowledge of the "saros cycle" (a cycle of 18 years, 10 or 11 days and 1/3 of a day that separates two successive sun or moon eclipses of similar features), inherited by the Greeks from the Babylonians, favored important theoretical developments in the West. The purpose of this paper is to present, briefly and schematically, a) the ancient theorizations on the causes of eclipses, b) the range of predictability of eclipses in Antiquity, c) the warnings –transmitted by some classical writers– to those who want to observe in naked eyes a solar eclipse, and d) the popular beliefs and social practices upon the occurrence of unexpected eclipses in ancient times.

**Keywords:** eclipses – astronomy – saros – cosmology[1]

El estudio de los eclipses en la Antigüedad ha despertado mucho mayor interés entre los científicos que entre los filólogos clásicos, especialmente a lo largo de los dos últimos siglos, cuando las descripciones de unos pocos eclipses, principalmente de sol, consagrados por la tradición astronómica como *eclipses históricos*, han sido consideradas científicamente como datos observacionales[2].

Los eclipses históricos, contrastados con modelos teóricos de la Astronomía, han llevado a su vez a reformulaciones serias de teorías científicas vigentes, en especial de la interacción entre la luna y la tierra, incluyendo aspectos tales como la aceleración secular aparente de la luna y la velocidad de rotación de nuestro planeta, cuya desaceleración pudo, gracias a esos relatos, ser en parte comprendida y explicada.

---

[1] El presente artículo completa y amplía el trabajo preliminar previo "Teorizaciones, predictibilidad, y creencias sobre los eclipses en la Antigüedad", presentado por los autores en el XXI Simposio Nacional de Estudios Clásicos, Santa Fe, Universidad Nacional del Litoral, septiembre de 2010.

[2] Los eclipses históricos son los descritos por fuentes históricas confiables –en su mayoría grecolatinas clásicas o tardoantiguas– cuyas fechas son compatibles con cálculos astronómicos retrospectivos. Se considera eclipses históricos a los siguientes: eclipse del canon asirio (15 de junio del 763 a.C.); eclipse de Asurbanipal (27 de junio del 661 a.C.); eclipse de Arquíloco (6 de abril del 648 a.C.); eclipse de Tales (28 de mayo del 585 a.C. –hay densas discusiones sobre este eclipse–); eclipse de Píndaro (30 de abril del 463 a.C.); eclipse de Tucídides (3 de agosto del 431 a.C.); eclipse de Agatocles (15 de agosto del 310 a.C.); eclipse de Hiparco (20 de noviembre del 129 a.C.); eclipse de Flegón (24 de noviembre del 29 d.C.); eclipse de Plutarco (20 de marzo del 71 d.C.); eclipse de Teón (16 de junio del 364 d.C. –aquí también quedan dudas–). Couderc, 1969: 128-134.

La tierra se traslada alrededor del sol en un lapso de 365,25 días, y lo hace en una órbita elíptica muy próxima a una circunferencia, donde el sol se ubica en uno sus focos. La luna, por su parte, cumple su órbita, también elíptica, alrededor de la tierra en aproximadamente unos 27,32 días. Pero esta órbita está inclinada con respecto a la de la tierra alrededor del sol (esta última, llamada la eclíptica) en unos 5,1º. Durante su movimiento orbital, entonces, la luna intersecta al plano de la eclíptica en dos puntos opuestos, llamados nodos. Sólo cuando la luna se ubica en alguno de esos nodos *pueden* producirse los eclipses, pues son las únicas ocasiones en las que tierra, sol y luna pueden hallarse casi perfectamente alineados, aproximadamente cada 5,8 meses sinódicos.

Pero la orientación de la órbita de la luna no es constante en el tiempo, sino que *precesa* (esto es, la ortogonal al plano de la órbita lunar cambia su dirección en el espacio, como si se tratase del eje de un trompo que pierde velocidad de giro). Y esto es debido en parte a la interacción de nuestro satélite con el sol y con la tierra. Como consecuencia, la línea imaginaria que une los nodos también cambia y se va corriendo lentamente por el plano de la eclíptica en sentido retrógrado (contrario al orden de los signos zodiacales) unos 19,3º por año.

Esta breve reseña ilustra una parte de la complejidad que involucra la explicación (y más aún, la predicción) de los eclipses. Por ejemplo, el *ciclo de saros* (ciclo que predice la recurrencia de eclipses similares) fue el resultado de la cuidadosa observación y registro de los eclipses llevado a cabo por la astronomía mesopotámica desde el siglo VIII a.C.[3]. Se trata de un ciclo de 6585,32 días –o bien 18 años, 10 (u 11) días y 1/3 de día, si dentro del ciclo se incluyen cinco (o cuatro) años bisiestos– que establece el período de tiempo que media entre dos eclipses totales de sol o de luna sucesivos con similares características. Para que uno de estos eventos suceda deben darse tres condiciones importantes: tratarse de luna nueva o llena (eclipse de sol o de luna, respectivamente), la luna debe ubicarse en uno de los dos nodos de su órbita (alineación cuasiperfecta entre los tres cuerpos) y, por último, la distancia entre la luna y la tierra debe ser la misma que en el eclipse recurrente del ciclo anterior (por ejemplo, para que el diámetro de la cobertura del disco solar por la silueta de la luna y la sombra consecuente, en el caso de un eclipse de sol, sean las mismas). El resultado es que la luna se hallará en la misma fase, se ubicará en igual nodo y a la misma distancia de nuestro planeta, sólo cada 6585,32 días.

---

[3] Steele, 2000: 422-424; 431-433.

La descripción y la predicción de eclipses ocupó a numerosos pensadores antiguos y tardoantiguos: mientras algunos elaboraron curiosas especulaciones sobre sus causas, otros lograron predecir con parámetros racionales la *posibilidad de un eclipse*. Y todos, de un modo u otro, favorecieron la comprensión de estos acontecimientos astronómicos, concretada definitivamente por la Europa moderna, heredera del mundo clásico. En efecto, las sucintas descripciones de eclipses –por poner algunos ejemplos– de Arquíloco, Píndaro o Tucídides (entre otros) han aportado datos decisivos para el esclarecimiento de problemas astronómicos impensados en la Antigüedad. El núcleo de su aporte surge de un rasgo común a todos los eclipses históricos, y es que para los estudiosos del siglo XIX aquellos eclipses coincidían en su *fecha* con la predicción teórica de eclipses, pero no en la *hora*, puesto que ocurrían, hasta posteriores correcciones, aproximadamente unas 3 horas antes (es decir, unos 45º de longitud terrestre hacia el Este) respecto de lo que el cálculo y las tablas lunares reclamaban. Como resultado de ello, la comunidad científica ensayó un conjunto de correcciones a la velocidad de rotación de la tierra, que debió ser concebida en un proceso de desaceleración como resultado de la fricción *tidal* o de mareas que se produce en la región de contacto entre los océanos y la corteza terrestre, la cual causa que la velocidad de rotación disminuya lentamente y el día, por tanto, se alargue a razón de –en promedio– un par de milisegundos por siglo[4]. Mediante dicha fórmula resultó posible la armonización –no sin dificultades originadas en otros factores, especialmente las perturbaciones gravitacionales producidas por otros cuerpos del sistema solar, especialmente Júpiter y Saturno, sobre los cuerpos astronómicos involucrados– del grueso de los eclipses históricos con la Teoría Lunar debidamente reformulada y corregida.

---

[4] La fricción contra el suelo de las cuencas oceánicas por las mareas actúa como un débil freno que ralentiza la rotación de la tierra. Recientes estimaciones de la desaceleración de la velocidad de rotación terrestre arroja el siguiente valor para la duración del día: 21,9 horas hace 620 millones de años (y, con grandes incertezas, aproximadamente 20,9 horas hace 900 millones de años y entre 17,1 y 18,9 horas hace 2.500 millones de años). Véanse Williams, 1997: 421-424; Couderc, 1969: 124-127.

Estrabón (ca. 63 a.C.-19 a 24 d.C.) en su *Geografía*, 1, 1, 12 define a los eclipses como συγκρίσεις ἡλίου καὶ σελήνης, es decir, como combinaciones o composiciones del sol y la luna [con la tierra]. Una apropiada definición, por cierto, que sigue siendo la usada mayormente en la actualidad. Estas alineaciones de los tres astros son de diversos tipos. Los eclipses de luna pueden ser totales o parciales, y en ambos casos pueden ser vistos por todos los habitantes de la tierra vueltos hacia la luna en ese momento (un 50% de la superficie terrestre en cada instante), momento por lo demás siempre nocturno, por tratarse de un fenómeno necesariamente de luna llena, con el sol y la luna en oposición. Son fenómenos interesantes, en la medida en que la luna toma un color rojizo oscuro, y porque la sombra esférica de la tierra se proyecta sobre la superficie lunar, ilustrando al ojo atento el tamaño aparente del cono de sombra terrestre, cuyo diámetro sobre la luna es de aproximadamente tres diámetros lunares, por lo que una sección del borde del cono se ve durante el evento perfectamente dibujada sobre Selene o Luna-Febe-Delia (tales los nombres que griegos y latinos daban al cuerpo lunar). Teniendo presente que la luna se desplaza en sentido directo sobre su órbita a razón de 12° en 24 horas y que el diámetro lunar es de aproximadamente 0,5°, la duración de un eclipse lunar es cercana a las tres horas, con al menos dos horas, en el caso de los eclipses plenos, de la fase de *totalidad*. Los eclipses de luna son fenómenos llamativos, pero salvo por las escasas situaciones en las que la luna desaparece completamente (como combinación del eclipse con fenómenos atmosféricos) su inspección no reviste mayor dramaticidad, ni comporta, desde luego, peligro alguno para la vista humana.

Otra cosa son los eclipses de sol, los cuales son básicamente de tres tipos: parciales, totales o anulares. Los eclipses parciales de sol suelen pasar inadvertidos para el observador desatento, y el amante de los eclipses debe cuidarse seriamente de no mirarlos a simple vista, por más alto que sea el porcentual de ocultación solar. Los eclipses anulares ocurren en virtud de que la distancia tierra-luna es levemente variable: cuando la luna se encuentra transitando su apogeo (máxima distancia a la tierra), su tamaño aparente es levemente más pequeño que el del sol, por lo que sin llegar a cubrirlo completamente deja un pequeño aro de luz solar visible. Durante estos eclipses la oscuridad sobre la superficie terrestre, si bien es manifiesta, no es total. Finalmente, los eclipses que pueden llegar a producir un aterrador efecto desde el punto de vista psíquico para quienes no están prevenidos, y aun temor e inquietud para los advertidos, son los eclipses totales de sol. Estos comienzan súbitamente con un sorprendente crepúsculo en horas en que no corresponde tal fenómeno, cuando la oscuridad se incrementa dramáticamente, hasta que en medio del día se produce la noche. Aparecen algunas estrellas prominentes y desciende inmediatamente la temperatura en forma drástica, provocando una alteración profunda en el comportamiento de los animales (incluido el hombre).

Concluimos esta introducción –necesaria para sentar las bases astronómicas de los fenómenos que nos convocan– puntualizando el propósito de nuestro trabajo. Es nuestro interés discutir sucintamente cuatro aspectos relevantes vinculados a los fenómenos eclípticos en la Antigüedad, a saber: a) las teorizaciones sobre las causas de los eclipses, b) la predictibilidad de estos singulares eventos astronómicos, c) las recomendaciones para la observación de eclipses de sol y, finalmente, d) algunas creencias y prácticas sociales manifestadas en ocasión de los eclipses. En las próximas secciones discutiremos estos aspectos relevantes para la explicación de los eclipses en el pasado y citaremos opiniones y testimonios de diversos pensadores y filósofos naturales de la cultura grecolatina clásica y tardoantigua.

**Teorizaciones sobre las causas de los eclipses**

Las teorizaciones antiguas grecorromanas sobre las causas de los eclipses revelan en general una actitud científica y son dignas de ser incorporadas con legitimidad a la historia de la ciencia. Los modelos surgidos, desde luego, involucran siempre aspectos cosmológicos que exceden la relación entre los tres astros (Gea, Helio, Selene), implicando consideraciones generales sobre el sistema solar y la naturaleza de los cuerpos celestes. Estos modelos revelan la consolidación de un marco explicativo, que es por lo demás el actualmente vigente, al que denominaremos aquí *modelo úmbrico*, por tener a las *sombras* como su concepto central. Reseñamos brevemente a continuación los principales modelos teóricos distinguibles en la Antigüedad[5].

*Modelo obturatorio*: El modelo obturatorio fue defendido por Anaximandro, y requiere un compromiso teórico con la comprensión del sol y la luna como anillos huecos que encierran fuego y que poseen en su centro una abertura circular, por la que emana la luz que nos llega habitualmente de dichos astros[6]. Anaximandro explica las mutantes fases de la luna, al igual que los eclipses, recurriendo a este modelo. Concibe pues a los cambios aspectuales de los astros como resultado de *obturaciones temporarias* de dichas bocas circulares, explicadas por fenómenos atmosféricos intrínsecos al sol y a la luna. Los testimonios más relevantes sobre estas ideas que han llegado a nosotros los debemos a Hipólito de Roma, Aecio y Simplicio, quienes manifiestan que Anaximandro tuvo en su vida una auténtica preocupación por explicar los eclipses y dan cuenta en forma sintética del modelo explicativo de dichos eventos propuesto por el cosmólogo presocrático. Simplicio (s. VI d.C.), en su *Comentario al De caelo [de Aristóteles]*, 471, 1 (correspondiente a *De caelo* II, 10, 291a), sostiene que Anaximandro pudo conocer las distancias al sol y a la luna gracias a su estudio de los eclipses (DK 12 A 19, LFP I 161), al tiempo que Hipólito de Roma (s. III d.C.), en su *Refutación de todas las herejías*, I, 6, 4 (DK 12 A 11, LFP I 165) señala que, según Anaximandro:

> …τὰ δὲ ἄστρα γίνεσθαι κύκλον πυρός, ἀποκριθέντα τοῦ κατὰ τὸν κόσμον πυρός, περιληφθέντα δ' ὑπὸ ἀέρος. ἐκπνοὰς δ' ὑπάρξαι πόρους τινὰς αὐλώδεις, καθ' οὓς φαίνεται τὰ ἄστρα· διὸ καὶ ἐπιφρασσομένων τῶν ἐκπνοῶν τὰς ἐκλείψεις γίνεσθαι.

---

[5] Sin llegar a alcanzar valor teórico, vale la pena tener presente al recorrer los tres modelos propuestos a continuación (obturatorio, friccional, úmbrico) que en el Fragmento 122 (West) de Arquíloco hay también una alusión a un eclipse de sol, el cual es atribuido, ya en clave religiosa, ya en clave poética, a la acción de Zeus.
[6] Kahn, 1960: 86-87.

> …los astros se generan como un círculo de fuego, separándose del fuego del mundo, circundado cada uno por aire. Hay orificios, conductos en forma de flautas, a través de los cuales se muestran los astros, por lo cual, cuando los orificios son obstruidos, se producen los eclipses[7].

Dicho testimonio coincide con el provisto por Aecio (s. I-II d.C.) en sus *Placita philosophorum* (compiladas posteriormente en los s. IV-V d.C.), quien atribuye a Anaximandro la idea de que el "eclipse de sol se produce al obstruirse la abertura por la que se exhala el fuego" (*Placita* II 24 2, DK 12 A 21, LFP I 167), mientras que "la luna se eclipsa al obstruirse la abertura de la rueda" (*Placita* II 29 I, DK 12 A 22, LFP I 168).

*Modelo friccional*: Denominamos así al que habría sido sostenido por Demócrito de Abdera (460 a.C.-ca. 370 a.C.), y que propone que la invisibilidad del sol o de la luna sería causada por un fenómeno ocasional de *descenso* de sus respectivas órbitas. Dichos movimientos, atribuidos por el vulgo a la acción de brujas o magos, eran denominados también, según el Escolio a Apolonio de Rodas III, 553 (DK 68 B 161, LFP III 554), "descendimientos", a los cuales se atribuía el oscurecimiento del cuerpo celeste eclipsado[8]. La idea principal de este modelo es la siguiente: el *rozamiento* o *fricción* de los cuerpos celestes contra el éter (tomamos aquí la licencia de utilizar el concepto aristotélico, que es el que la tradición doxográfica que transmite la doctrina democritea tiene en mente) se reduciría por la propia disminución de la *velocidad lineal* de los cuerpos (pues cuanto más cerca del centro orbitan, menor es su rapidez), ocasionando con ello que merme su brillo. Otros testimonios que permiten sostener esta idea son provistos por

---

[7] VV.AA., 2000: 161.

[8] Respecto de la producción humana de eclipses mediante hechicería refiere tangencialmente Platón en *Gorgias* 513a que las brujas tesalias sufrían, como resultado del esfuerzo por hacer descender a estas luminarias, serios daños físicos, al parecer quemaduras y parálisis en las piernas (véase nota de F. García Yagüe en Platón, 1980: *ad locum*). Por otra parte, Estrepsíades, el limitado alumno del *pensadero* de Sócrates en *Las nubes*, también propone pedir a una maga de Tesalia que le baje la luna para entonces encerrarla en una caja y evitar que pase el tiempo, medido esencialmente por los ciclos lunares, ahorrándose así la necesidad de pagar sus deudas (*Las nubes*, v. 750). Plinio el Viejo, en *Naturalis historia*, II, 9, también comenta que el vulgo consideraba a los eclipses como resultado de la acción de brujas, y que por esa razón quienes eran sorprendidos por un eclipse producían ruidos con cuanto hubiera a mano para evitarlos.

Lucrecio y por Diógenes Laercio[9]. La misma idea, aunque en tono jocoso, nos la ofrece Aristófanes (*Las nubes*, vv. 584 y ss.), cuando el Corifeo anuncia que en una ocasión en que la ciudad estaba eligiendo para el cargo de estratego a un personaje que disgustaba a Aristófanes, en el cielo "la luna abandonó su curso y el sol retrayendo con prontitud en sí mismo su mecha se negaba a brillar"[10].

*Modelo úmbrico*: Denominamos de esta forma a la consideración de los fenómenos de eclipse como resultado de la interposición de la luna o de la tierra en la línea que une al sol y el cuerpo eclipsado. Según este modelo, los eclipses son simplemente el resultado óptico causado por la sombra de un astro (luna o tierra) sobre el otro. Este modelo ha sido el dominante en la Antigüedad, siendo numerosísimos los autores que lo propusieron. Mencionamos sólo unos pocos: Aristóteles, en *De caelo* II, 11, 291b24, afirma que los eclipses son producto de la sombra del cuerpo eclipsante y destaca que los eclipses de luna son la principal prueba de la esfericidad de la tierra, en la medida en que el borde circular de la sombra de la tierra se evidencia con gran claridad sobre la superficie lunar. Igualmente Gémino (s. I a.C.), en su *Introducción a los Fenómenos [de Arato]*, XI, propugna esta idea, e Higino (64 a.C.-17 d.C.) se suma a la larga lista de autores que proponen este modelo, ofreciendo en su *Astronomía*, IV, 14, 3, la siguiente explicación de los eclipses:

> Lunae autem eclipsis sic evenit, cum prope dimensione sit luna, cum abierit sol sub terram, dumtaxat hoc modo, ut per mediam terram si quid directum traieceris, contingere possit solem sub terra, lunam autem supra terram; quod cum ita evenit, necesse est solis radios propter magnitudinem terrae ita esse dimissos, ut lumen eius, quo luna lucet, non possit ad eam pervenire, et ita existimatur fieri eclipsis lunae.

---

[9] Lucrecio, *De rerum natura*, V, vv. 621-636 (DK 68 A 88, LFP III 554) y Diógenes Laercio, *Vidas de los filósofos ilustres*, IX, 33 (DK 67 A 1, LFP III 543). Por lo demás, ya en otro orden de cosas, un curioso texto de Diógenes Laercio (*Vidas*, I. Prólogo), que propone a Hefesto como padre, entre los egipcios, de la Filosofía, sugiere que éste vivió 48.863 años antes que Alejandro Magno, y que en ese intervalo hubo 373 eclipses solares y 832 eclipses lunares, es decir, un eclipse solar cada 131 años y un eclipse lunar cada 58,7 años. Los datos provistos por Diógenes son curiosos, y causan perplejidad, pues no coinciden en absoluto con la frecuencia de los eclipses (por cierto ya conocida en su tiempo).
[10] Aristófanes, 1999: 65.

Por su parte, un eclipse de luna tiene lugar cuando ésta [la luna] se encuentra alineada con el sol, opuesto a ella por debajo de la tierra. De este modo, si se traza una línea recta por el centro de la tierra, puede tocar por debajo el sol y por encima la luna. Cuando esto ocurre, los rayos del sol, debido a la magnitud de la tierra, forzosamente son desviados, de tal modo que la luz gracias a la que la luna brilla, no le puede llegar. Es así como se cree que se produce un eclipse de luna[11].

Igualmente Plinio el Viejo (62-113 d.C.), en su *Historia natural*, II, 7, 47 avala este modelo afirmando que:

Quippe manifestum est solem interuentu lunae occultari lunamque terrae obiectu ac uices reddi, eosdem solis radios luna interpositu suo auferente terrae terraque lunae. Hac subeunte repentinas obduci tenebras rursumque illius umbra sidus hebetari. Neque aliud esse noctem quam terrae umbram, figuram autem umbrae similem metae ac turbini inuerso, quando mucrone tantum ingruat neque lunae excedat altitudinem, quoniam nullum aliud sidus eodem modo obscuretur et talis figura semper mucrone deficiat.

Efectivamente, es cierto que el sol se eclipsa por la interposición de la luna, la luna por la intercalación de la tierra, y ambos eclipses son equivalentes, ya que con su respectiva interposición la luna quita a la tierra (y la tierra a la luna) los mismos rayos de sol; también, que al introducirse la luna, se originan inmediatamente las tinieblas, y a su vez, el tal astro se oscurece por la sombra de la tierra, asimismo, que la noche no es otra cosa que la sombra de la tierra, pues la forma de la sombra es similar a un cono o a una peonza con el pico hacia arriba, porque cae sobre la luna exclusivamente por su punta y no excede su altura, siendo así que ningún otro astro se

---

[11] Higino, 2008: 347.

oscurece del mismo modo y que una figura como ésa siempre termina en punta[12].

**Predictibilidad de los eclipses en la Antigüedad**

Los círculos científicos del mundo clásico y helenístico mostraron gran interés en el problema de la predicción de los eclipses. Sin embargo una predictibilidad exitosa de los fenómenos eclípticos no fue posible sino tras la divulgación de la teoría gravitatoria newtoniana, y resultados confiables han sido alcanzados recién durante el siglo XVIII, y mejorados sustancialmente durante el siglo XIX[13]. Sólo muy recientemente las predicciones de eclipses han alcanzado gran precisión, con márgenes de error ínfimos, y actualmente se destacan entre los especialistas los cálculos de Jean Meeus y Fred Espenak, autores del *Catalog of Solar Eclipse Saros Series – Five Millennium Canon of Solar Eclipses*, disponible como publicación electrónica de la NASA, en los que se basa gran parte de la comunidad científica.

El modelo cosmológico dominante en la Antigüedad, que concibe al universo como el espacio (interno) limitado por la esfera celeste (o esfera de las estrellas fijas), tuvo su punto fuerte precisamente en el trazo de una serie de curvas esféricas (ecuador celeste, eclíptica, coluros tropicales o solsticiales, órbita lunar, etc.) que permitieron –en virtud de su simplicidad y mensurabilidad (angular)– un conocimiento empírico-estadístico del ciclo de los eclipses. Tal conocimiento fue, en rigor, heredado por el mundo grecorromano de las investigaciones mesopotámicas, las que ya desde el siglo VIII a.C. adquirieron una comprensión primaria de estos fenómenos, que alcanzó cierta precisión para el caso de los eclipses de luna (no así para los de sol).

El método usado por la astronomía mesopotámica consistía en lo siguiente (lo describimos en términos modernos): distinguiendo el *mes sidéreo* (27,32 días), del *mes sinódico* (29,53 días), del *mes draconítico* (27,21 días), y siendo el *mes sidéreo* el período que tarda la luna en cruzar dos veces seguidas la longitud celeste de una determinada estrella, el *mes sinódico* el intervalo entre dos conjunciones luna-sol, y el *mes draconítico* el intervalo entre dos pasos sucesivos de la luna por un mismo nodo, se obtienen los elementos teóricos mínimos con los que es posible comprender la *posibilidad* de un eclipse, y establecer las reglas de su predicción. Estos tres conceptos son accesibles a la observación sistemática, y su distinción remite inmediatamente al ciclo de saros, pues en 6585 días y 1/3 de día se cumplen simultáneamente 223 meses sinódicos, 242 meses draconíticos y (aunque esto último no lo sabían los antiguos) 239 meses anomalísticos[14]. Esa coincidencia hace que los eclipses se vean con características muy similares tras la ocurrencia de dicho período.

---
[12] Plinio el Viejo, 1995: 354.
[13] Airy, 1853: 179-182.

Expresado más simplemente, sólo cuando la sombra de la tierra en oposición o la del sol en conjunción (mes sinódico) están cerca de un nodo (mes draconítico) será posible un eclipse. Esto es equivalente a afirmar que los eclipses sólo son posibles cuando la latitud eclíptica de la luna es muy cercana a 0º en el momento de la conjunción o de la oposición, hecho que coincide además, necesariamente, con el cumplimiento simultáneo de un número entero de meses sinódicos y de meses draconíticos. El modo más confiable de predecir la posibilidad de un eclipse consiste entonces en calcular la latitud eclíptica de la luna en el momento de las *sizigias*[15], y establecer –según una razón dictada por la experiencia de observación y favorecida por la regularidad de los movimientos orbitales de la luna– que aquellas sizigias en las que la latitud de la luna sea cercana a 0º indicarán la *posibilidad de un eclipse*. Este era, en efecto, el método utilizado por la astronomía matemática babilonia durante la dominación helenística del período seléucida (312-63 a.C.)[16]. Y puesto que la latitud de la luna es fácilmente mensurable con un simple sextante o astrolabio, los astrónomos mesopotámicos llegaron a advertir que cada aproximadamente 5,8 meses sinódicos se daban las dos condiciones mínimas requeridas para un eclipse, a saber, que la luna se encontrase en una latitud cercana a 0º y que el nodo se hallase cerca de la sizigia.

---

[14] El *mes anomalístico* es el tiempo que media entre dos sucesivos pasajes de la luna por su perigeo (téngase presente que la órbita lunar es elíptica, siendo el perigeo su punto más cercano a la tierra).

[15] El término *sizigia* indica simplemente en forma genérica las *conjunciones* y las *oposiciones* (la palabra griega $\sigma\upsilon\zeta\upsilon\gamma\iota\alpha$ significa *unión, combinación, conjunción*). Los eclipses ocurren cuando la sizigia está lo suficientemente cerca del nodo de la luna.

[16] Steele, 2000: 423.

Así pues, la astronomía babilonia temprana alcanzó una sencilla regla que establece, sobre la base de las mencionadas distinciones, que un eclipse puede ocurrir sólo cinco o seis meses después de un eclipse anterior. El uso de esta regla tiene sin embargo una dificultad, en virtud de la incertidumbre respecto de cuándo es conveniente intercalar el intervalo de 5 meses, pues en cada saros (223 lunaciones) hay 38 posibilidades de eclipses, 33 de las cuales están separadas por seis meses mientras que las restantes 5 cada 5 meses. Es así que distribuyendo parejamente las posibilidades de que el siguiente eclipse ocurra 5 meses después de otro eclipse a lo largo de un saros (6585,32 días), se obtienen diversas secuencias posibles, del tipo 8-7-8-7-8, siendo 7 u 8 la cantidad de posibilidades de eclipses sucesivas separadas entre sí por seis meses tras las cuales se inserta la *posibilidad de un eclipse* tras una brecha de 5 meses respecto de la posibilidad de eclipse anterior. Dicho más sencillamente, a 8 posibilidades de eclipses que se suceden con una brecha de 6 meses le sigue una posibilidad de eclipse separada por una brecha de 5 meses, y luego siete posibilidades de eclipse con brechas de 6 meses, y así sucesivamente. Este patrón varía según cada ciclo de saros, pues estos pueden comenzar por una brecha de cinco o de seis meses respecto del último eclipse del ciclo de saros anterior, por lo que resultan las siguientes cinco series como la totalidad de las posibilidades de distribución de las brechas de 5 meses: 8-7-8-7-8; 7-8-7-8-8; 8-7-8-8-7; 7-8-8-7-8; 8-8-7-8-7.

El registro sistemático de las posibilidades de eclipse y de los eclipses observables permitió así a la astronomía mesopotámica el conocimiento general de este ciclo de 6585 días y 1/3 de día –el 1/3 de día indica que el eclipse recurrente de un saros a otro ocurrirá corrido en 8 horas, es decir, desplazado unos 120º hacia otra comarca geográfica ubicada al Oeste–[17]. En el mundo grecorromano, el conocimiento de este ciclo de 223 lunaciones está atestiguado por numerosos textos y aun por restos materiales, los cuales señalan en su conjunto la central consideración que dieron al saros los filósofos naturales antiguos. Estrabón (*Geografía*, I, 1) informa que el gran astrónomo de la tardoantigüedad, Hiparco (ca. 190-120 a.C.), solía afirmar que nadie puede enseñar geografía o astronomía, ni distinguir latitudes de ciudades, si no tiene un adecuado conocimiento del ciclo de los eclipses, mientras que Marco Tulio Cicerón (106-43 a.C.) en *De divinatione* II, 1, señala que los eclipses fueron predichos muchos años antes de su época por hombres versados en matemáticas, en virtud de que los movimientos de la luna son perfectamente regulares, por lo que es posible calcular cuándo la luna ingresa en el cono de sombra de la tierra.

---

[17] El ciclo conocido como *exeligmos* (correspondiente a tres ciclos de saros) retorna el eclipse a aproximadamente la misma longitud terrestre. Es menester distinguir (véase Steele, 2000: 424) entre el *ciclo de saros* (223 lunaciones), la *serie de saros* (colección de posibilidades de eclipses de semejantes características separadas por diversos ciclos de saros), y el *esquema de saros* (particular distribución de las posibilidades de eclipses dentro de un ciclo de saros o, dicho de otro modo, la disposición de las posibilidades de eclipse cada cinco meses dentro del ciclo, que puede variar según que esta brecha de 5 meses se de en el primer eclipse del ciclo de saros en cuestión o en alguno de los siguientes hasta el octavo –máxima distancia posible entre dos brechas de 5 meses sucesivas–).

Un ejemplo material del conocimiento del ciclo es el famoso mecanismo de Anticitera, complejo artilugio mecánico descubierto entre los restos de un naufragio cerca de la isla griega de Anticitera, vecina a Creta, y datado hacia el 87 a.C., que posee entre sus ábacos uno que registra el ciclo de saros y que permite a sus usuarios –sin duda avezados astrónomos y navegantes– conocer la posibilidad cierta de un eclipse con sólo disponer el mecanismo en estación[18].

Asimismo Plinio el Viejo (23-79 d.C.), quien dedica el capítulo II, 10 de su *Historia natural* al análisis de la recurrencia de los eclipses de sol y de luna, sintetiza el resultado del método mesopotámico descrito, tomándolo de Hiparco:

> Defectus CCXXIII mensibus redire in suos orbes certum est, solis defectus non nisi nouissima primare fieri luna, quod uocant coitum, lunae autem non nisi plena, semperque citra quam proxime fuerint; omnibus autem annis fieri utriusque sideris defectus statis diebus horisque sub terra nec tamen, cum superne fiant, ubique cerni, aliquando propter nubila, saepius globo terrae obstante conuexitatibus mundi. Intra ducentos annos Hipparchi sagacitate compertum est et lunae defectum aliquando quinto mense a priore fieri, solis uero septimo.

---
[18] Freeth, Jones, Steele & Bitsakis, 2008: 614-617.

Es cosa comprobada que los eclipses se repiten en los respectivos globos a los doscientos veintitrés meses, que el eclipse de sol sólo ocurre en el último cuarto de luna o en el primero (que es lo que llaman la conjunción); en cambio, el de luna sólo en plenilunio y siempre más acá de donde se produjo la última vez y, además, que todos los años se suceden los eclipses de ambos astros, en días y horas fijos, bajo la tierra; cuando se producen por encima de ella, no son visibles desde todas partes, a veces por las nubes y más a menudo por el obstáculo del globo terráqueo a causa de la forma abovedada del universo. Hace doscientos años se supo, gracias a la sagacidad de Hiparco, que el eclipse de luna ocurría a veces a los cinco meses del anterior, y el de sol, en cambio, a los siete[19].

La predicción de posibilidades de eclipses de luna alcanzó mediante este método un grado considerable de efectividad. En cambio, el caso de los eclipses totales de sol, como sabemos, es más complejo, pues la sombra del sol dibuja sobre la superficie terrestre un círculo de apenas unos 200 km de diámetro, u óvalos algo más grandes, que se desplazan en una franja muy estrecha del globo a gran velocidad (de hecho, a velocidades supersónicas). Esa franja de tierra ensombrecida se "corre" de un saros a otro, por lo que aun teniendo la vivencia efectiva de un eclipse y el conocimiento del ciclo, hubiese resultado imposible para un pensador antiguo predecir con certeza la siguiente ocurrencia de un eclipse de sol para una determinada comarca.

Existe vasta literatura sobre el famoso eclipse de Tales de Mileto (ca. 639-ca. 547 a.C.), y sobre la afirmación de Heródoto (485-452 a.C.) (*Los nueve libros de la historia*, I, 74) de que Tales habría predicho un eclipse de sol, pero la comunidad astronómica tiende a ver el hecho como una fantasiosa construcción doxográfica (con algo inclusive de hagiográfica) aun cuando tibiamente se admite la posibilidad de que, en caso de haber sabido Tales del eclipse de sol del 18 de mayo del 603 a.C., hubiese podido anticipar, gracias a su eventual conocimiento del saros –del que no hay pruebas–, el eclipse solar del 28 de mayo del 585 a.C.[20]. Los eclipses totales de sol ocurren aproximadamente cada 370 años sobre un mismo punto de la tierra, por lo que su ocurrencia definitivamente no pudo ser predicha por los antiguos[21].

---
[19] Plinio el Viejo, 1995: 358.
[20] Couderc, 1969: 131.

**Precauciones para la observación de eclipses de sol**

El peligro que representa la observación directa (a ojo desnudo) de los eclipses de sol fue bien conocido por los filósofos antiguos. Hoy sabemos que mirar directamente al sol, sin un filtro adecuado, es altamente peligroso: dado que la retina no está inervada, la destrucción de las células de su parte central no ocasiona ningún dolor. El observador desprevenido puede sufrir una alteración definitiva de la vista, incluso un daño como ceguera irreversible, sin ningún síntoma o molestia que le sirvan de advertencia. Platón (ca. 427-347 a.C.), en *Fedón* 99d, pone en boca de Sócrates estas palabras:

> Ἔδοξε τοίνυν μοι, ἦ δ' ὅς, μετὰ ταῦτα, ἐπειδὴ ἀπειρήκη τὰ ὄντα σκοπῶν, δεῖν εὐλαβηθῆναι μὴ πάθοιμι ὅπερ οἱ τὸν ἥλιον ἐκλείποντα θεωροῦντες καὶ σκοπούμενοι πάσχουσιν· διαφθείρονται γάρ που ἔνιοι τὰ ὄμματα, ἐὰν μὴ ἐν ὕδατι ἤ τινι τοιούτῳ σκοπῶνται τὴν εἰκόνα αὐτοῦ. τοιοῦτόν τι καὶ ἐγὼ διενοήθην, καὶ ἔδεισα μὴ παντάπασι τὴν ψυχὴν τυφλωθείην βλέπων πρὸς τὰ πράγματα τοῖς ὄμμασι καὶ ἑκάστῃ τῶν αἰσθήσεων ἐπιχειρῶν ἅπτεσθαι αὐτῶν. ἔδοξε δή μοι χρῆναι εἰς τοὺς λόγους καταφυγόντα ἐν ἐκείνοις σκοπεῖν τῶν ὄντων τὴν ἀλήθειαν.

> Después de esto tuve para mí, desde que fracasé en el estudio de las cosas [inteligibles], que era necesario cuidarme de que no me sucediera como a aquellos que miran y observan un eclipse de sol: a veces algunos pierden la vista por no observar en el agua o en algún otro modo la imagen del sol. Yo pensé algo por el estilo, y temía así quedar completamente ciego del alma, al mirar las cosas [inteligibles] con los ojos y esforzarme en ponerme en contacto con ellas por medio de cada uno de los sentidos. Juzgué, pues, que era necesario refugiarme en las proposiciones y buscar en ellas la verdad de las cosas[22].

---

[21] Magnasco & Baikouzis, 2008: 8823. Estos autores establecieron –sobre la base de investigaciones de Schoch y Neugebauer de comienzos del siglo XX– una fecha probable para los episodios descritos en la *Odisea*, proponiendo al 16 de abril de 1178 a.C. como fecha altamente probable para la alusión a un eclipse de sol mencionado en el canto XX, v. 356 y ss., en el marco del episodio conocido como la "Profecía de Teoclímeno", que anuncia cual portento la inminente matanza de los pretendientes a manos de Odiseo y su hijo Telémaco.

Aunque en este texto Platón sólo se refiere a lo inapropiado que resulta intentar captar la realidad inteligible de las Ideas mediante los sentidos del cuerpo, del pasaje surge claramente que los eclipses de sol, especialmente luego de terminada la fase de totalidad, eran considerados en los círculos letrados de la Grecia clásica eventos a ser contemplados con sumo cuidado y munidos en todo caso de elementos apropiados. Un interesante testimonio de las *Cuestiones naturales*, I, 11, 3-12, 1 de Séneca (4 a.C.-65 d.C.) revela, en la misma dirección, los cuidados esperables para la contemplación segura de un eclipse:

> Quotiens defectionem solis uolumus deprehendere, ponimus pelues, quas aut oleo aut pice implemus, quia pinguis umor minus facile turbatur et ideo quas recipit imagines seruat; apparere autem imagines non possunt nisi in liquido et immoto. Tunc solemus notare, quemadmodum luna soli se opponat et illum tanto maiorem obiecto corpore abscondat, modo ex parte, si ita competit, ut in latus eius incurreret, modo totum; haec dicitur perfecta defectio, quae stellas quoque ostendit et intercipit lucem, tunc scilicet cum uterque orbis sub eodem libramento stetit.
>
> Cuando queremos observar un eclipse de sol, colocamos en el suelo recipientes llenos de aceite o de pez, porque un líquido denso no se agita con facilidad y retiene mejor las imágenes que reproduce. Las imágenes no pueden reflejarse sino en líquido tranquilo e inmóvil. Entonces observamos cómo se interpone la luna entre nosotros y el sol; cómo este astro, siendo mucho más pequeño que aquél, colocándose delante, le oculta en parte unas veces, si solamente le opone un lado, y otras por completo. Llámase eclipse total el que hace aparecer las estrellas interceptando la luz, y tiene lugar cuando el centro de los dos astros se encuentra en la misma línea con relación a nosotros[23].

---
[22] Platón, 1983: 183-184.

**Creencias y prácticas sociales ante los eclipses**

La incertidumbre natural sobre las verdaderas causas de los eclipses y el miedo que generaban en los pueblos, que veían en estos fenómenos la ruptura del orden y de la armonía natural de los cielos, se ven reflejados en las creencias de los antiguos sobre la relación entre los eclipses y el mundo humano. Dichas creencias se encuentran enraizadas en prácticas divinatorias y mágicas de cuño mítico-religioso, las cuales dominaron el imaginario antiguo, especialmente la cultura popular. Esas prácticas incluían la interpretación de sucesos naturales o cósmicos en forzada relación con situaciones humanas, por lo que a menudo los portentos naturales eran interpretados como causados por la propia acción de los hombres o pueblos.

Así, al menos, lo manifiesta un interesante texto de Tucídides (460-396 a.C.), en el que –narrando episodios de la Guerra del Peloponeso– afirma que durante esta terrible contienda fueron más frecuentes los terremotos y los eclipses que lo que eran habitualmente (*La Guerra del Peloponeso*, I, 23)[24]. Asimismo, Virgilio (70-19 a.C.), en sus *Geórgicas* I, vv. 464-468, al detallar los pronósticos del tiempo, incluye la relevancia del estudio de la luna y del sol, y agrega a propósito que Febo fue el anunciador de la guerra civil y de los desastres que siguieron a la muerte de Cayo Julio César el 15 de marzo del año 44 a.C., pues:

---

[23] Séneca, 1884: 264-265.
[24] Una idea similar aparece en el famoso Peán 9 de Píndaro (Frag. 52K Maehler = A1 Rutherford), que describe un eclipse de sol. El poeta se pregunta de qué guerra traerá el eclipse señal, o si estará anunciando una nevada, o el anegamiento de la llanura por el avance del mar o acaso una sedición, entre otras calamidades naturales y sociales. Bellamente, concluye Píndaro el discurso afirmando que nada es digno de ser temido si se lo ha de sufrir junto a todos los otros miembros de su comunidad.

> … Ille etiam caecos instare tumultus
> saepe monet fraudemque et operta tumescere bella;
> ille etiam exstincto miseratus Caesare Romam,
> cum caput obscura nitidum ferrugine texit
> impiaque aeternam timuerunt saecula noctem.

El Sol nos avisa también muchas veces de la inminencia de perturbaciones secretas, de que se están fraguando traiciones y guerras en la sombra. Él fue el que, compadecido de Roma a la muerte de César, cubrió su brillante cabeza de obscura herrumbre, y las generaciones impías temieron una noche eterna[25].

Por otra parte, diversos relatos de escritores antiguos sobre reacciones espontáneas ante eclipses resultan elocuentes para ilustrar el temor que en el bajo pueblo provocaban estos fenómenos atípicos. Baste recordar dos episodios, recogidos por Plutarco y Tácito, autores coetáneos de la segunda mitad del siglo I d.C., uno del ámbito helenístico y otro del latino.

Plutarco (50-120 d.C.), en su vida de *Aemilius Paullus*, 17, 3, describe las reacciones contrapuestas que provocó un eclipse de luna entre las tropas macedónicas y las tropas romanas, y aún dentro de las romanas, la diferente reacción e interpretación que recibió el portento entre los soldados y los oficiales. El episodio se enmarca en la campaña de Paulo Emilio (229-160 a.C.), general romano de la época republicana que ejerció dos veces el consulado y que llevó adelante con éxito la Tercera Guerra Macedónica, deponiendo y haciendo prisionero al rey Perseo de Macedonia. El eclipse tuvo lugar en la noche del 21 al 22 de junio del 168 a.C., pocas horas antes de la famosa batalla de Pidna, que consolidó el dominio romano en Macedonia. Según el relato de Plutarco, tras el control del Monte Olimpo por la columna romana dirigida por Nasica, el ejército de Paulo Emilio divisa al enemigo poco antes del atardecer, y los jóvenes oficiales ansiosos por entrar en batalla presionan en busca de la orden de ataque. Pero Emilio sugiere cautela, considerando la inconveniencia de lanzar a hombres fatigados por el trajín de la jornada al campo de batalla contra un ejército fresco y descansado, y ordena armar campamento. Ya habiendo oscurecido y antes de medianoche tiene lugar un eclipse de luna por nadie previsto, y se suscitan en ambos bandos una serie de escenas que describe Plutarco elocuentemente:

> Ἐπεὶ δὲ νὺξ γεγόνει καὶ μετὰ δεῖπνον ἐτράποντο πρὸς ὕπνον καὶ ἀνάπαυσιν, αἰφνίδιον ἡ σελήνη πλήρης οὖσα καὶ μετέωρος ἐμελαίνετο, καὶ τοῦ φωτὸς ἀπολείποντος αὐτὴν χρόας ἀμείψασα παντοδαπὰς ἠφανίσθη. τῶν δὲ Ῥωμαίων, ὥσπερ ἐστὶ νενομισμένον, χαλκοῦ τε πατάγοις ἀνακαλουμένων τὸ φῶς αὐτῆς καὶ πυρὰ πολλὰ δαλοῖς καὶ δᾳσὶν ἀνεχόντων πρὸς τὸν οὐρανόν, οὐδὲν ὅμοιον ἔπραττον οἱ Μακεδόνες, ἀλλὰ φρίκη καὶ θάμβος τὸ στρατόπεδον κατεῖχε, καὶ λόγος ἡσυχῇ διὰ <τῶν> πολλῶν ἐχώρει, βασιλέως τὸ φάσμα σημαίνειν ἔκλειψιν. ὁ δ' Αἰμίλιος οὐκ ἦν μὲν ἀνήκοος οὐδ' ἄπειρος παντάπασι τῶν ἐκλειπτικῶν ἀνωμαλιῶν, αἳ τὴν σελήνην περιφερομένην εἰς τὸ σκίασμα τῆς γῆς ἐμβάλλουσι τεταγμέναις περιόδοις καὶ ἀποκρύπτουσιν, ἄχρι οὗ παρελθοῦσα τὴν ἐπισκοτουμένην χώραν πάλιν ἀναλάμψῃ πρὸς τὸν ἥλιον·

---

[25] Virgilio, 1994: 109. En otro registro, Virgilio alude a los eclipses de sol como fenómenos dignos de ser celebrados. En *Eneida*, II, vv. 740-755, en ocasión de la recepción brindada por la reina cartaginesa Dido a los troyanos, el bardo Iopas (discípulo del mismísimo Atlas) canta –mientras Dido se enamora de Eneas– al camino errante de la luna y a los eclipses del sol (v. 743: *hic canit errantem lunam solisque labores*), en el marco de un rico discurso cosmológico. Asimismo, en *Geórgicas*, II, v. 475 y ss. Virgilio exalta el conocimiento de la naturaleza, expresado poéticamente mediante un llamado a las Musas, en tanto ellas pueden ilustrar sobre los caminos del cielo y las estrellas, los eclipses de sol, las fases de la luna, las causas de los temblores de tierra, etc.

Al hacerse de noche, y cuando después del rancho se iban a dormir y descansar, la luna, que estaba en su lleno y bien descubierta, empezó de pronto a ennegrecerse, y desfalleciendo su luz, habiendo cambiado diferentes colores, desapareció. Los romanos, como es de ceremonia, la imploraban para que les volviese su luz, con el ruido de los metales, y alzando al cielo muchas luces con tizones y hachas; mas los Macedonios a nada se movieron, sino que el terror y espanto se apoderó del campo, y entre muchos corrió secretamente la voz de que aquel prodigio significaba la destrucción de su rey. No era Emilio hombre enteramente nuevo y peregrino en las anomalías que los eclipses producen, los cuales a tiempos determinados hacen entrar la luna en la sombra de la tierra y la ocultan, hasta que pasando de la sombra vuelve otra vez a resplandecer con el sol[26].

Al parecer, los soldados macedonios intuyeron correctamente el mensaje del eclipse. Si bien la batalla duró una hora, la carnicería romana continuó hasta bien entrada la madrugada del 23 de junio, arrojando más de 20.000 cadáveres de guerreros macedonios sobre las colinas adyacentes a Pidna, y sumando otros 11.000 prisioneros, luego esclavizados y deportados[27].

El testimonio de Tácito (55-120 d.C.), en sus *Anales*, I, 28, aporta una descripción similar de la reacción de otra tropa romana ante un eclipse. El texto describe cómo Druso Julio César (13 a.C.-23 d.C.) supo aprovechar un eclipse de luna para aplacar un motín inminente en la Germania, instigado por los oficiales Percenio y Vibuleno. Tácito destaca la reacción de los soldados, quienes al ver la luna oscurecida, asustados y creyendo que se trataba de un mal presagio sobre sus planes sediciosos, produjeron inmediatamente estruendosos ruidos con metales, trompetas y cuernos, y se mostraban ora apenados, ora exaltados mientras la luna se opacaba o aclaraba. También señala que tal reacción se fundamentaba a su juicio en la ignorancia de los soldados, aludiendo tácitamente a que entre los hombres letrados no existía duda respecto de que los eclipses son fenómenos naturales explicables racionalmente. Por lo demás, según dicho relato, el propio Druso, aprovechando la ocasión, envió a su centurión y a sus oficiales a pasar revista a la tropa y a arengar a los soldados, los que –atemorizados por el eclipse– respondieron favorablemente renovando la confianza en su jefe.

---

[26] Plutarco, 1821, II: 160.
[27] Este relato de Plutarco que describe a la luna como cambiando de color hasta desaparecer completamente ($\pi\alpha\nu\tau o\delta\alpha\pi\grave{\alpha}\varsigma\ \mathring{\eta}\phi\alpha\nu\acute{\iota}\sigma\theta\eta$) es tomado como una de las pocas descripciones antiguas de un *eclipse oscuro*, nombre que reciben los eclipses de luna en los que ésta deja de verse completamente (ello ocurre muy ocasionalmente, siendo tales fenómenos ocasionados ya por la opacidad de la atmósfera terrestre, ya por menguas en la actividad de la superficie solar). P. J. Bicknell (1968, 22) corrigió –precisamente citando este texto de Plutarco– a William Guthrie (1962, II: 307), quien había afirmado que no existe registro de este tipo de eclipses de luna en la Antigüedad. Richard B. Stothers (1986, 96), por su parte, indica que Plutarco podría haberse basado en textos hoy perdidos del científico romano Cayo Sulpicio Galo o del historiador Polibio, ambos testigos oculares del eclipse, aunque también podría haber inferido la oscuridad absoluta del eclipse basándose en ideas de la astrología caldea –por él bien conocidas–, que asociaban la visibilidad del eclipse con la hora de su ocurrencia (el texto de Plutarco indica que fue tras la cena). En suma, Stothers asume que esta descripción de Plutarco no debe tomarse como dato observacional confiable.

Un episodio parecido es narrado también por Plutarco, en su *Vida de Dión*, 24. Dión era hermano político y consejero de Dionisio, el famoso tirano de Siracusa que convocó a Platón también como consejero, y a quien –tras disgustarse con sus doctrinas– mandó primero matar, y luego, reviendo la condena, ordenó vender como esclavo en Egina (sus amigos debieron unirse para comprarlo y liberarlo). Tras la muerte del tirano, Dión hizo un viaje a Atenas, y armó un pequeño ejército para tomar el poder en Siracusa, luchando contra Dionisio el Joven, heredero del tirano. Ya en campaña, ofreció un banquete a sus soldados, y por la noche se produjo un eclipse de luna. Él no se preocupó porque conocía (por su formación en la Academia de Atenas) las causas del eclipse y la regularidad de sus ocurrencias, pero Miltas, su adivino de cabecera, ante el temor de los soldados, interpretó el portento como señal de que eclipsaba el reinado de Dionisio el Joven, y aprovechó la ocasión para arengar a los soldados.

Por otra parte, no sólo los soldados en situación de conflicto, cuando sus pensamientos se hacen permeables a todo portento ominoso, sino también los astrólogos-astrónomos romanos consideraban a los eclipses como portadores de peligros. Con claridad lo manifiesta Manilio en su *Astronomica,* tal el nombre del famoso poema didáctico (IV, vv. 839-848), donde considera que los *signos zodiacales* pierden parte de su potencia en ocasión de los eclipses, quedando como heridos en ciertos meses, expresándose así sobre la causa de dicho fenómeno:

> Causa patet, quod, Luna quibus defecit in astris
> orba sui fratris noctisque inmersa tenebris,
> cum medius Phoebi radios intercipit orbis
> nec trahit adsuetum quo fulget Delia lumen,
> haec quoque signa suo pariter cum sidere languent
> incuruata simul solitoque exempta uigori
> et uelut elatam Phoeben in funere lugent.

La razón es clara porque, al sufrir eclipse la luna en algunos signos, privada de su hermano [Febo, el sol] y sumergida en las tinieblas de la noche, cuando la tierra, situada en medio, intercepta los rayos de Febo y Delia [la luna] no atrae la luz con la que brilla normalmente, también esos signos [zodiacales, vg. su eficacia] languidecen junto con su planeta [la luna] y, arqueándose al mismo tiempo y perdiendo su acostumbrado poder, lloran a Febe en sus exequias, como si estuviera en su funeral[28].

Esta personificación de los astros pertenecientes al fondo estrellado del cielo, también tocó al sol y a la luna, que fueron a menudo caracterizados en la literatura clásica como seres vivientes, pasibles de sufrir enfermedades y hasta la muerte. Los eclipses eran en este sentido concebidos como la manifestación de la impotencia o debilidad, temporaria quizás, de la luna o del sol. Lucrecio (99-ca. 55 a.C.) así lo narra en su poema *De rerum natura* (V, vv. 751-770): "¿Y no puede el sol mismo eclipsarse y perder en cierta hora también su brillo, que recobra al punto que atravesó por medio de los aires regiones enemigas de sus llamas y le precisan a extinguir sus fuegos?". O bien, cuando al cuerpo brillante de la luna tócale transitar entre aires densos y hostiles, "¿no puede ella lentamente eclipsarse en cierta parte del Mundo, atravesando por parajes capaces de apagar sus mismos fuegos?"[29].

Plinio el Viejo, décadas más tarde, también escribió sobre las creencias de los hombres en lo que respecta a la supuesta muerte de los astros durante los eclipses, y celebra en su *Naturalis historia*, II, 9, la liberación de la ignorancia que emana de la explicación racional de los fenómenos, exaltando las ideas de Tales de Mileto e Hiparco sobre la mecánica de los eclipses, y ensalzando la obra de estos grandes hombres por haber descubierto las leyes con que se mueven los cuerpos divinos. Con ello, pretende Plinio, se ha liberado a la empobrecida mente de los hombres del temor que los eclipses comportaban, y de las vanas creencias sobre los –supuestos– portentos nefastos que anunciaban, que incluían la idea de que las estrellas mismas serían destruidas por su ocurrencia. A todos estos temores solía responder el pueblo espontáneamente, ante la ocurrencia de un eclipse, mediante sonidos colectivos sincronizados emitidos con el propio cuerpo, con cualquier objeto disponible a mano o eventualmente con instrumentos de percusión.

---
[28] Manilio, 1996: 220. Los corchetes son nuestros.
[29] Lucrecio, 1997: 322.

Una idea semejante es expresada por Severino Boecio (480-ca. 526), quien afirma en *La consolación de la Filosofía*, libro IV, metro 5, que al eclipsarse Febe "la ignorancia del pueblo turba a las naciones y fatigan los címbalos con sus constantes golpes"[30], al tiempo que propone (libro I, prosa 4) a la Astronomía, en tanto muestra las *vias siderum* (ie. en tanto enseña los caminos de los astros), como una actividad fundamental para la Filosofía, único camino de vida auténticamente liberador del hombre que conduce a la verdadera felicidad o sumo bien. Su reflexión es propicia para sintetizar la justa celebración que merecen los geómetras y astrónomos antiguos por su contribución al esclarecimiento de las leyes del cielo, pues su ingente esfuerzo, visto a la distancia, constituye un perenne monumento a la agudeza y a la pujanza humanas en busca del conocimiento pleno y articulado de las causas de los fenómenos, tarea en la que tan profundamente se ocupó –como lo muestra el tratamiento de los eclipses– la cultura clásica.

**Agradecimientos**

---

[30] Boecio, 1997: 266. Tito Livio, en su *Historia de Roma desde su fundación* (*Ab Urbe condita libri*), 26, 5, 9, también señala esta práctica.

**FUENTES**

## BIBLIOGRAFÍA